\documentclass[aps,showpacs,twocolumn,longbibliography]{revtex4-1}
\usepackage[utf8]{inputenc}
\usepackage{amsfonts}
\usepackage{amssymb}
\usepackage{amsmath}
\usepackage{graphicx}
\usepackage{epsfig}
\usepackage{subfigure}
\usepackage{appendix}
\usepackage{color}
\usepackage{hyperref}
\usepackage{clrscode3e}

\hypersetup{hypertex=true,
colorlinks=true,
linkcolor=blue,
anchorcolor=blue,
urlcolor=blue,
citecolor=blue}

\begin{document}

\title{Dissipation-induced bound states as a two-level system}
\author{H. P. Zhang}
\author{Z. Song}
\email{songtc@nankai.edu.cn}

\begin{abstract}
Potential wells are employed to constrain quantum particles into forming
discrete energy levels, acting as artificial few-level systems. In contrast,
an anti-parity-time ($\mathcal{PT}$) symmetric system can have a single pair
of real energy levels, while all the remaining levels are unstable due to
the negative imaginary part of the energy. In this work, we investigate the
formation of bound states in a tight-binding chain induced by a harmonic
imaginary potential. Exact solutions show that the real parts of energy
levels are equidistant, while the imaginary parts are semi-negative definite
and equidistant. This allows for the formation of an effective two-level
system. For a given initial state with a wide range of profiles, the evolved
state always converges to a superposition of two stable eigenstates. In
addition, these two states are orthogonal under the Dirac inner product and
can be mutually switched by applying a $\pi$ pulse of a linear field. Our
finding provides an alternative method for fabricating quantum devices
through dissipation.
\end{abstract}

\affiliation{School of Physics, Nankai University, Tianjin 300071, China}
\maketitle

\section{Introduction}

\label{Introduction} The bound state concept pervades various branches of
physics, including optics and condensed matter. A bound state refers to a
particle confined within a localized region by a real-valued potential,
often due to interactions with other particles, within the framework of the
Hartree-Fock approximation. Throughout the field of physics, phenomena that
are stable or in equilibrium can be elucidated by the concept of bound
states, which applies to both quantum and classical systems. Experimentally,
bound states can be engineered by introducing artificial defects into
photonic crystals. These defects, arranged in an array, form what are known
as coupled-resonator optical waveguides (CROWs), which enable nearly
lossless guidance and manipulation of wave packets, including their bending 
\cite%
{Johnson2001,John1987,Yablonovitch1987,Skorobogatiy2005,Engelen2008,Hughes2005,Kuramochi2005,LeThomas2008,Mazoyer2009,Mazoyer2010,O’Faolain2007}%
.

In contemporary physics, the presence of a complex potential is now
permissible, as non-Hermitian quantum mechanics has become a versatile
framework for developing functional devices that operate within the
non-Hermitian domain. {The fundamental mechanism underlying non-Hermitian
systems with parity-time ($\mathcal{PT}$){\ symmetry }hinges on
the concept of an imaginary potential} \cite%
{Bender1998,Dorey2001,Mostafazadeh2002,Znojil1999,Jones2005,ElGanainy2007,Musslimani2008,Makris2008,Joglekar2010,Scott2011,Chong2010,Jing2014}%
.  This concept has been both theoretically explored and experimentally
realized \cite%
{Guo2009,Rueter2010,Wan2011,Sun2014,Feng2012,Peng2014,Chang2014,Feng2014,Hodaei2014,Wimmer2015}%
, serving as an essential building block for constructing such systems.

A non-Hermitian Hamiltonian exhibits unique characteristics that set it
apart from its Hermitian counterpart in three distinct aspects. (i)
Asymmetry hopping term can result in skin effect, a typical bound state \cite%
{Zhang2013,Kunst2018,Yao2018,Gong2018,Jin2019}. (ii) Adding a
pseudo-Hermitian term to a non-Hermitian Hamiltonian consistently reduces
the level spacing. Conversely, a nontrivial Hermitian perturbation
invariably results in level repulsion. The phenomenon of shrinking level
spacing is analogous to the impact of a real potential on a quantum system 
\cite{Zhang2019}. (iii) In a system with energy levels featuring
semi-negative imaginary components, only those eigenstates with real-valued
energies are considered stable.

\begin{figure}[tbh]
\centering
\includegraphics[width=0.45\textwidth]{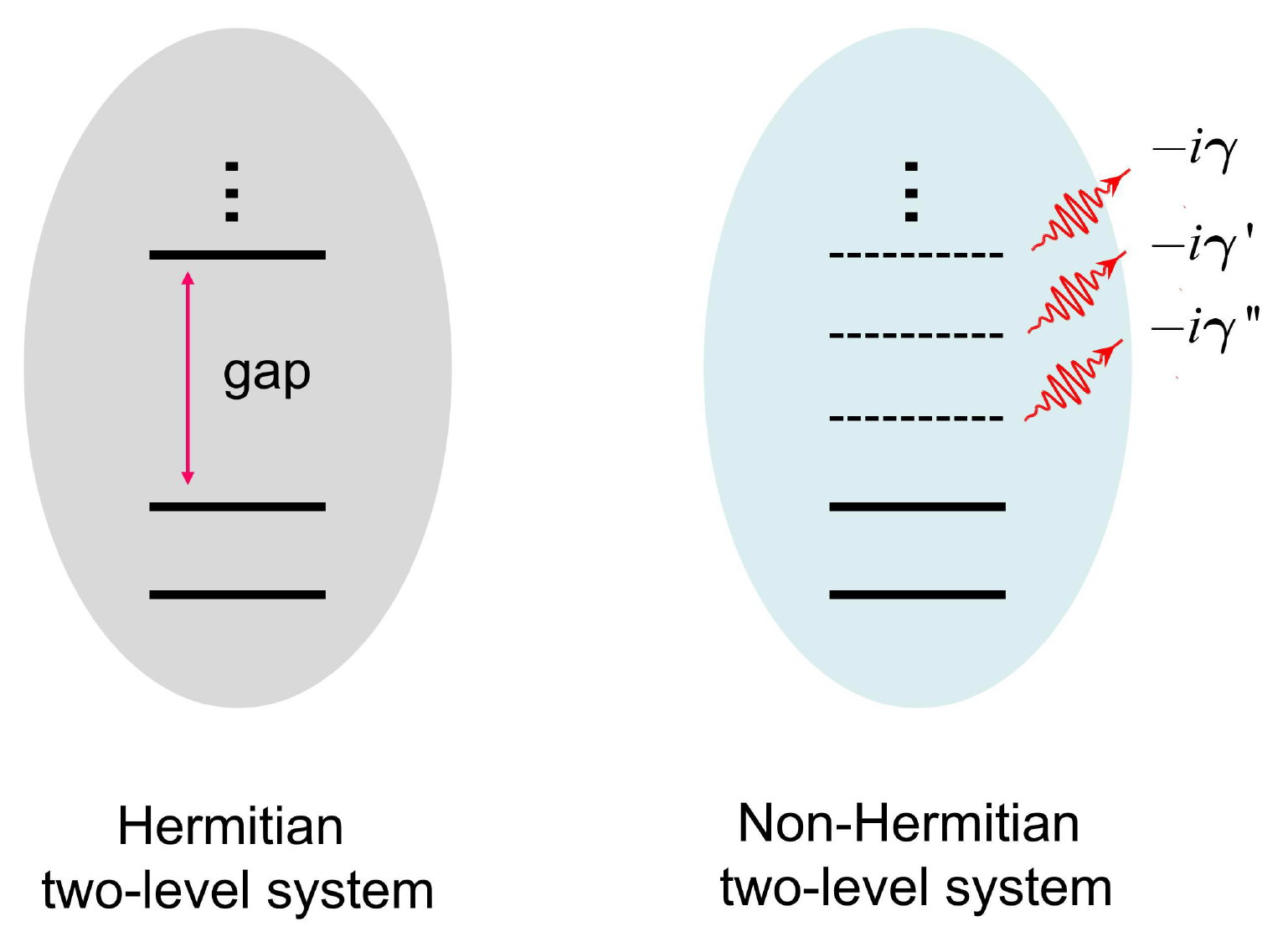}
\caption{Hermitian two-level system vs. non-Hermitian two-level system. (a)
A Hermitian two-level system can be achieved in a many-body system when the
gap between the first-excited state energy and the higher energy levels is
sufficiently large. The time evolution preserves the Dirac probability for
any initial state. (b) A non-Hermitian two-level system can be achieved in a
many-body system when all the eigenstates decay with different rates $%
\protect\gamma $, $\protect\gamma ^{\prime }$, $\protect\gamma ^{\prime
\prime }$ and so on, except the two lower levels.\ In general, the Dirac
probability for any initial state is periodic due to the biorthogonality of
two-level states. Nevertheless, the proposed non-Hermitian two-level system
can act as a Hermitian one.}
\label{fig1}
\end{figure}

In this work, we explore the formation of two-level system through the
mechanism associated with the characteristic mentioned as feature (iii) (see
fig.\ \ref{fig1}). We investigate the formation of bound states in a
tight-binding chain induced by a harmonic imaginary potential. It is an anti-%
$\mathcal{PT}$ symmetric system. Unlike $\mathcal{PT}$-symmetry accompanied
by real eigenvalues, an anti-$\mathcal{PT}$ symmetric system can have a
single pair of real energy levels, while all the remaining levels are
unstable due to the negative imaginary part of the energy. The phenomenon of
dissipation gives rise to bound states that can be effectively described as
a two-level system. We consider a tight-binding chain induced by a harmonic
imaginary potential to demonstrate this feature. Exact solutions show that
the real parts of energy levels are equidistant, while the imaginary parts
are semi-negative definite and equidistant. This allows for the formation of
an effective two-level system. For a given initial state with a wide range
of profiles, the evolved state always converges to a superposition of two
stable eigenstates. We also find that these two states are orthogonal under
the Dirac inner product, and thus a superposition state obeys the Dirac
probability-preserving dynamics. Therefore, although the two-level system is
engineered by non-Hermitian terms, it acts as a conditional Hermitian
device. In addition, it is shown that two states can be mutually switched by
applying a $\pi $ pulse of a linear field. We demonstrate the conclusions by
numerical simulations. Our finding provides an alternative method for
fabricating quantum devices through dissipation.

This paper is organized as follows. In Sec. \ref{Hamiltonian and symmetry},
we introduce the non-Hermitian model and present approximate solutions for
low energy levels. In Sec. \ref{Two-level dynamics} and \ref{Transition
between two levels}, we show that two-level quantum states can be
dynamically prepared and operated. Finally, our conclusion is given in Sec. %
\ref{Conclusion and Discussion}.

\begin{figure*}[tbh]
\centering
\includegraphics[width=0.85\textwidth]{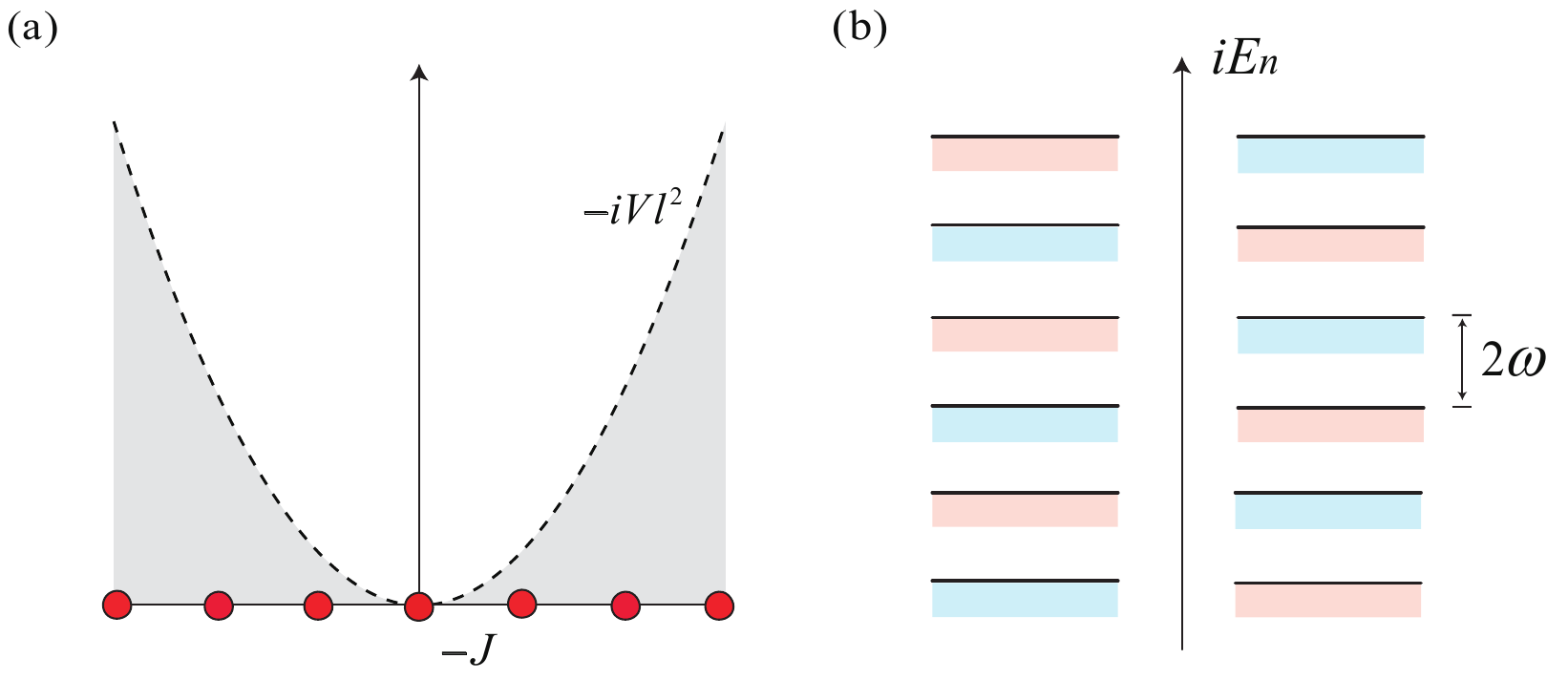}
\caption{(a) Schematic illustrations of the Hamiltonian in Eq.\ (\protect\ref%
{e1}), which represents a tight-binding chain with adjacent hopping
strengths is $-J$ and on-site imaginary potential $-iVl^{2}$ (a constant $i%
\protect\omega $ is omitted). The anti-$\mathcal{PT}$ symmetry of the system
ensures the following characteristics for the energy level structure. (b)
Energy level structure diagrams for Hamiltonian of (a). $E_{n}$ stands for
eigenenergy of the Hamiltonian in Eq.\ (\protect\ref{e1}). The solid line
represents the real part of $iE_{n}$ with isoenergetic distance $2\protect%
\omega $. The red and blue blocks represent the positive imaginary part and
negative imaginary part of $iE_{n}$, respectively. It can be seen that $%
iE_{n}$ is composed of conjugate pair energy levels due to the pesudo
Hermiticity of the Hamiltonian $iH$.}
\label{fig2}
\end{figure*}

\section{Hamiltonian and symmetry}

\label{Hamiltonian and symmetry}

We start with the tight-binding model with the Hamiltonian%
\begin{equation}
H=-J\sum\limits_{l=-\infty }^{+\infty }(|l\rangle \langle l+1|+|l+1\rangle
\langle l|)-iV\sum\limits_{l=-\infty }^{+\infty }l^{2}|l\rangle \langle
l|\,+i\omega ,  \label{e1}
\end{equation}%
where $|l\rangle $ denotes a site state describing the Wannier state
localized on the $l$th period of the potential. Here, $-J$ is the tunneling
strength, and $V$\ is the strength of the harmonic potential. The system is
schematically illustrated in fig.\ \ref{fig2}(a). In previous work, it has
been shown that the lower energy levels are equidistant when the system is
Hermitian by taking $V\rightarrow iV$ \cite{Shi2005,Tao2005}. In this work,
we take $J$, $V>0$, and constant $\omega =\sqrt{JV/2}$. We note that the
Hamiltonian has anti-$\mathcal{PT}$ symmetry, satisfying%
\begin{equation}
\mathcal{PT}H\left( \mathcal{PT}\right) ^{-1}=-H,
\end{equation}%
where the linear operator $\mathcal{P}$\ and antilinear operator $\mathcal{T}
$\ are defined as%
\begin{equation}
\mathcal{P}|l\rangle =(-1)^{l}|l\rangle ,\mathcal{T}i\mathcal{T}^{-1}=-i.
\end{equation}%
The eigenstate ${\left\vert \psi _{m}\right\rangle }$, satisfying the
Schrodinger equation 
\begin{equation}
{H\left\vert \psi _{m}\right\rangle =E_{m}\left\vert \psi _{m}\right\rangle ,%
}  \label{Sch eq}
\end{equation}%
\ can be written in the form 
\begin{equation}
{\left\vert \psi _{m}\right\rangle =\sum_{l}\psi _{m}(l)|l\rangle .}
\end{equation}%
According to the non-Hermitian quantum mechanics \cite{Scholtz1992}, $iH$ is
a pseudo-Hermitian operator, and then $i{E_{m}}$\ can be real or come in
complex conjugate pairs.

Specifically, we expect the solution for the wave function ${\psi _{m}(l)}$
to be connected to a continuous function, ${\psi _{m}(x)}$. In the
following, we aim to obtian the approximate expression of the function ${%
\psi _{m}(x)}$, under certain conditions. The Schrodinger equation (\ref{Sch
eq}) shows that function ${\psi _{m}(l)}$\ can be determined by the equation

\begin{equation}
\psi _{m}(l+1)+\psi _{m}(l-1)=\frac{i\omega -E_{m}-iVl^{2}}{J}\psi _{m}(l).
\label{e6}
\end{equation}%
We propose two types of functions $\psi _{m}^{\pm }(l)$\ as the Bethe ansatz
wave function, which are defined as%
\begin{equation}
\psi _{m}^{+}(l)=N_{m}\exp (-\frac{\alpha ^{2}l^{2}}{2})H_{m}(\alpha l),
\end{equation}%
and%
\begin{equation}
\psi _{m}^{-}(l)=\left( -1\right) ^{l}[\psi _{m}^{+}(l)]^{\ast },
\end{equation}%
with coefficients $\alpha =e^{-i\pi /8}(V/J)^{1/4}$ and $\left( N_{m}\right)
^{-2}=$ $\int_{-\infty }^{+\infty }H_{m}\left( \alpha x\right) H_{m}\left(
\alpha ^{\ast }x\right) $ $\times e^{-\func{Re}\alpha ^{2}x^{2}}dx$.
Obviously, $\psi _{m}^{\pm }(x)$\ are eigen functions of quantum oscillator
system, satisfying the differential equation%
\begin{equation}
\left[ \frac{\mathrm{d}^{2}}{\mathrm{d}x^{2}}-\alpha ^{4}x^{2}+\alpha
^{2}(2m+1)\right] \psi _{m}^{+}=0.  \label{e9}
\end{equation}%
with $m=0$, $1$, $2$,$...$. As known that function $\psi _{m}^{+}(x)$ is
smooth and slowly varying for small $m$ and $\alpha $ (or $V\ll J$), where $%
H_{m}\left( \alpha x\right) $\ is the Hermite polynomials. This allows the
approximation%
\begin{equation}
\psi _{m}^{+}(x+1)+\psi _{m}^{+}(x-1)\approx 2\psi _{m}^{+}(x)+\frac{\mathrm{%
d}^{2}}{\mathrm{d}x^{2}}\psi _{m}^{+}\left( x\right) ,
\end{equation}%
resulting in the eigen energy levels 
\begin{equation}
E_{m}^{+}=\left[ (2m+1)\omega -2J\right] -i2m\omega ,
\end{equation}%
{in association with} Eqs. (\ref{e6}) and (\ref{e9}). Similarly, we have 
\begin{equation}
E_{m}^{-}=\left[ 2J-(2m+1)\omega \right] -i2m\omega .
\end{equation}%
It accords with the prediction {with the relations} 
\begin{equation}
\mathcal{PT}\left\vert \psi _{m}^{+}\right\rangle =\left\vert \psi
_{m}^{-}\right\rangle ,
\end{equation}%
and%
\begin{equation}
\left( iE_{m}^{+}\right) =\left( iE_{m}^{-}\right) ^{\ast },
\end{equation}%
i.e., the real parts of the levels are paired with opposite signs. The
additional feature of the energy levels is that their imaginary parts are
semi-negative finite and equally distant, which is schematically illustrated
in fig.\ \ref{fig2}(b).

Remarkably, there are only two eigenstates have real energy, with wave
function and eigen energy

\begin{eqnarray}
\psi _{0}^{\pm }(l) &=&\left( \pm 1\right) ^{l}N_{0}e^{-\omega
l^{2}/(2J)}e^{\pm i\omega l^{2}/(2J)}, \\
E_{0}^{\pm } &=&\mp (2J-\omega) ,
\end{eqnarray}%
where the normalization factor $N_{0}=[V/(2J\pi ^{2})]^{1/8}$. All the rest
eigenstates decay as time increases. In this sense, such a system only
supports two stable eigenstates, the ground state 
\begin{equation}
\left\vert \text{g}\right\rangle =|\psi _{0}^{+}\rangle ,
\end{equation}%
and the excited state 
\begin{equation}
\left\vert \text{e}\right\rangle =|\psi _{0}^{-}\rangle .
\end{equation}

{In general, two different eigenstates are biorthogonal for
	non-Hermitian system, which satisfy the relationship, i.e., $\langle \varphi
	_{n}^{\sigma }|\psi _{m}^{\sigma ^{\prime }}\rangle =\delta _{nm}\delta
_{\sigma \sigma ^{\prime }}$\cite{Brody2013}. }It accords with the fact that we have
\begin{equation}
\langle \varphi _{0}^{\pm }|\psi _{0}^{\mp }\rangle =\sum_{l=-\infty
}^{+\infty }(-1)^{l}N_{0}^{2}e^{-\omega l^{2}/J}=0,
\end{equation}%
in the limit $\omega /J\rightarrow 0$ (or $\alpha \rightarrow 0$), where $%
|\varphi _{0}^{\pm }\rangle $\ satisfies 
\begin{equation}
{H}^{\ast }{|\varphi _{0}^{\pm }\rangle =(E_{0}^{\pm })^*|\varphi _{0}^{\pm
}\rangle .}
\end{equation}
Meanwhile, we also have%
\begin{equation}
\langle \text{e}|\text{g}\rangle =\sum_{l=-\infty }^{+\infty
}(-1)^{l}N_{0}^{2}e^{-(1-i)\omega l^{2}/J}=0,
\end{equation}%
in such a limit, which indicates that the Dirac orthogonality still holds.
Therefore, such a non-Hermitian system can be regarded as a Hermitian
two-level system.

\begin{figure*}[tbh]
\centering
\includegraphics[width=0.99\textwidth]{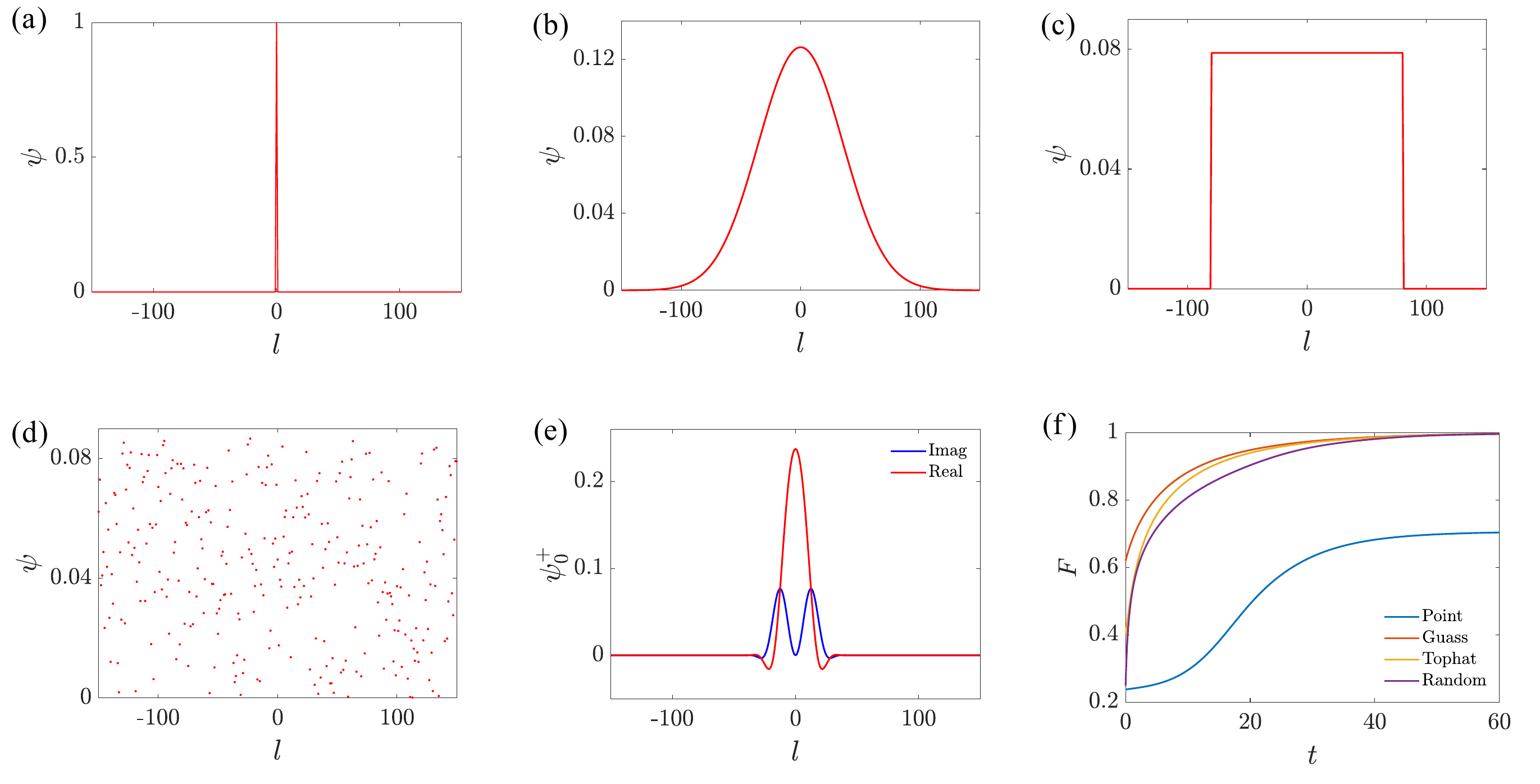}
\caption{Plots of the profiles of initial states and the fidelities of the
corresponding evolved states to the given target states. (a-d)\ Initial
states feature several typical amplitude distributions: point (delta
function), Gaussian, tophat, and random (stochastic) functions. (e) is the
plot of ground state of the non-Hermitian two-level system as the target
state. (f) is the plot of fidelity $F(t)$ defined in Eq.\ (\protect\ref{Ft}%
). We can see that the final states for (b), (c) and (d) are the same, the
ground state, while the superposition of ground state and excited state for
(a). The unit of time is $1/J$ and the system parameters are $J=1$, $%
V=2\times 10^{-4}$, and $\protect\omega =10^{-2}$.}
\label{fig3}
\end{figure*}

\begin{figure*}[tbh]
\centering
\includegraphics[width=0.97\linewidth]{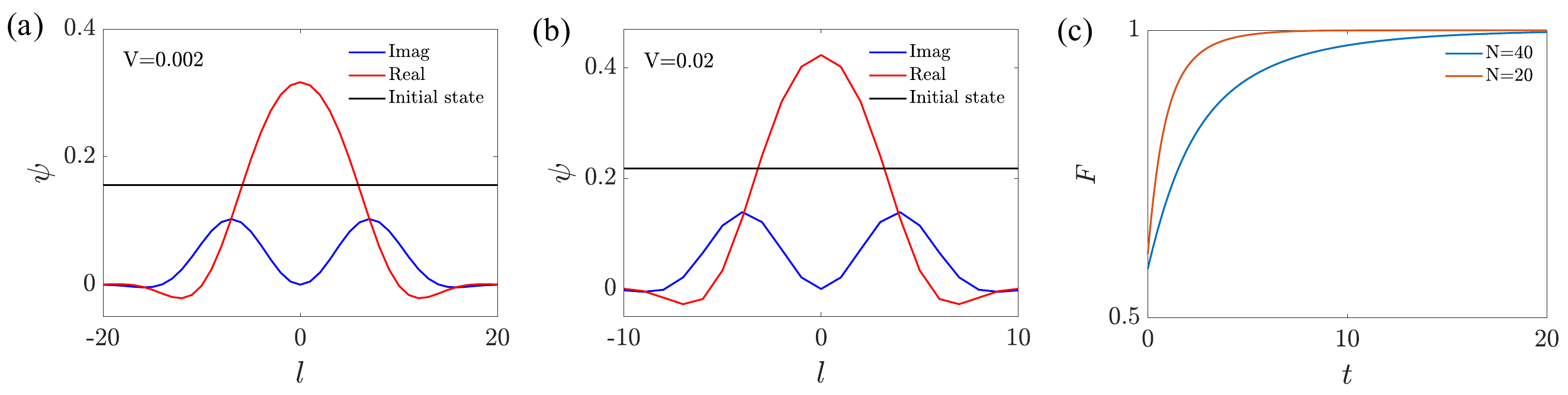}  
\caption{Plots of the profiles of ground states (target states)
under different system parameters and the fidelities of the corresponding
evolved states to the given target states. The system parameters are (a) $%
N=40$, $V=0.002$ and $J=1$; (b) $N=20$,$\ V=0.02$ and $J=1$. The red line
and blue line represent the real and imaginary parts of the ground state,
respectively, while the black line indicates the initial state. The plot of
fidelity $F(t)$ in (c) defined in Eq.\ \protect\ref{Ft} for (a) and (b) is
obtained by numerical simulation. It can be observed that increasing $V$
significantly shortens the time required for the fidelity to reach $1$.}
\label{fig4}
\end{figure*}

\begin{figure}[tbh]
\centering
\includegraphics[width=0.47\textwidth]{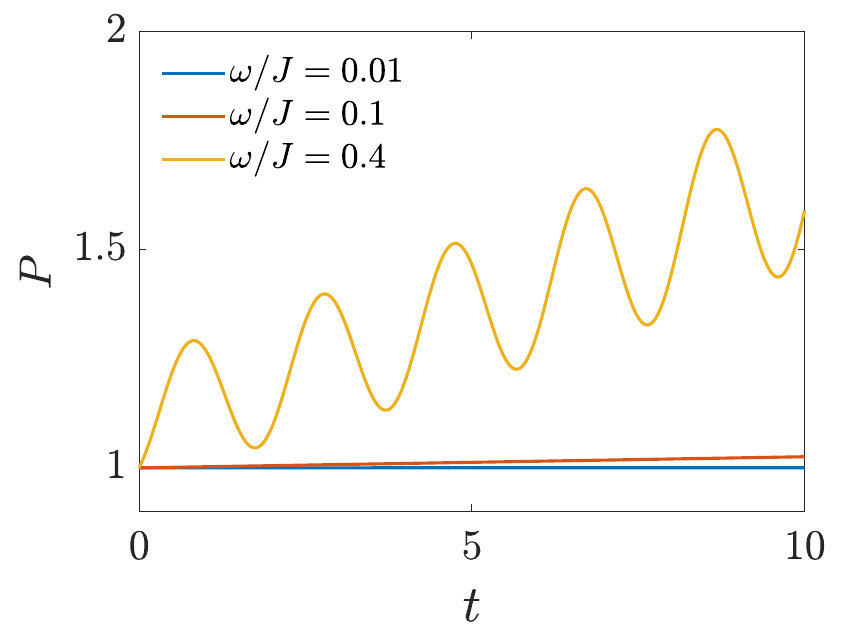}
\caption{Plots of $P(t)$ defined in Eq.\ (\protect\ref{e26}) for several
different $\protect\omega/J$. (a)(b)(c) respectively correspond to $\protect%
\omega/J=0.01$, $0.1$, $0.4$. It can be seen that when $\protect\omega/J$ is
small (<0.1), the Dirac probability is almost conserved, which indicates
that our approximation is appropriate. The unit of time is $1/J$ and the
system parameters are $J=1$.}
\label{fig5}
\end{figure}
\begin{figure*}[tbh]
\centering
\includegraphics[width=0.85\textwidth]{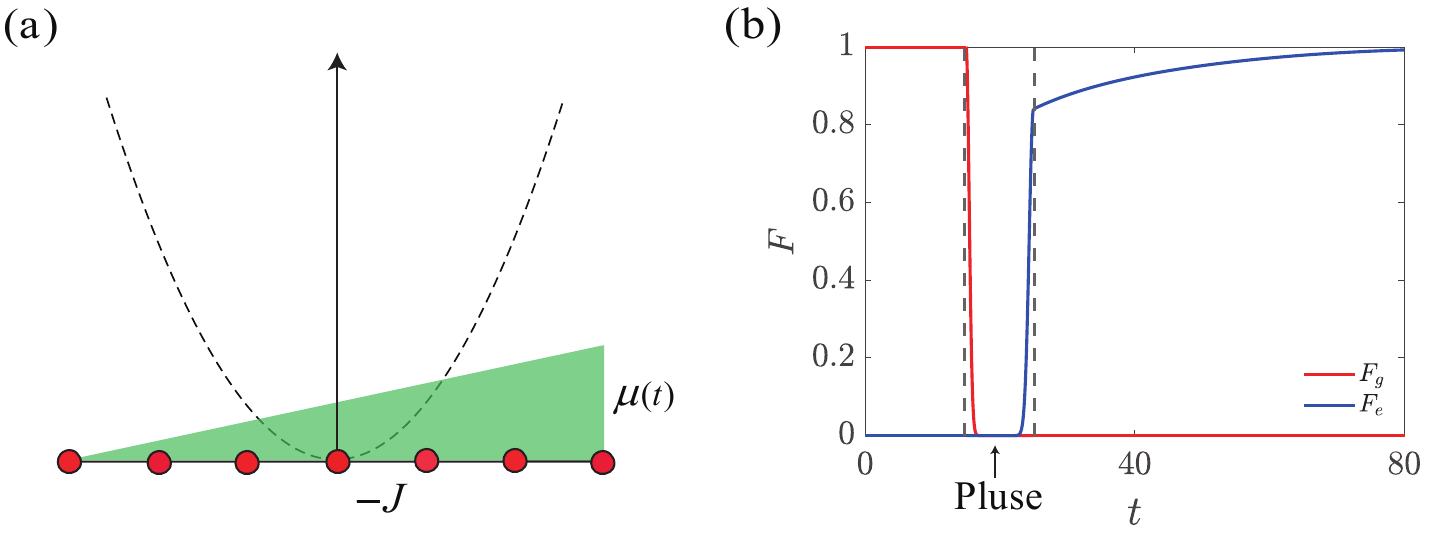}
\caption{ (a) Schematic illustrations of the Hamiltonian in Eq.\ (\protect
\ref{e28}), which represents adding the time varying skew potential pulse $%
\protect\mu(t)$ on the basis of the Hamiltonian in Eq.\ (\protect\ref{e1}).
(b) The plot of $F_g(t)$ and $F_e(t)$ defined in Eqs.\ (\protect\ref{Fg})
and (\protect\ref{Fe}) for initial state $\left\vert \text{g}\right\rangle$.
For the case with initial state $\left\vert \text{e}\right\rangle$, an
almost same figure is obtained by switching $F_g(t)$ with $F_e(t)$. It can
be seen that the conversion between the ground state and the excited state
can be realized by applying $\protect\pi$ pulse. The unit of time is $1/J$
and the system parameters are $J=1$, $V=2\times 10^{-4}$, and $\protect%
\omega =10^{-2}$.}
\label{fig6}
\end{figure*}

\section{Two-level dynamics}

\label{Two-level dynamics}

In quantum physics, two-level atoms are simple yet rich quantum systems that
are fundamental to our understanding of quantum mechanics and are essential
for the development of quantum technologies. The ability to control and
manipulate these systems is at the heart of many advances in quantum science.

In a Hermitian system, an effective two-level system can be formed when the
ground state and the excited state have relative large energy gap from the
high excited states. In contrast, a distinguishing characteristic of the
non-Hermitian two-level system is that the all other eigenstates are
unstable in dynamics or short lived (see fig.\ \ref{fig1}). This enables the
system to exhibit the following dynamical behaviors.

We start with the time evolution for an arbitrary initial state $|\Phi
(0)\rangle $, which can always be expressed as the form%
\begin{equation}
|\Phi (0)\rangle =\sum_{n}\left( c_{n}^{+}|\psi _{n}^{+}\rangle
+c_{n}^{-}|\psi _{n}^{-}\rangle \right) ,
\end{equation}%
in the framework of the above approximation approach. Thus, the time evolved
state is%
\begin{equation}
|\Phi (t)\rangle =\sum_{n,\pm }c_{n}^{\pm }e^{-iE_{n}^{\pm }t}|\psi
_{n}^{\pm }\rangle ,
\end{equation}%
which tends to%
\begin{equation}
|\Phi (t)\rangle =c_{0}^{+}e^{-iE_{0}^{+}t}\left\vert \text{g}\right\rangle
+c_{0}^{-}e^{-iE_{0}^{-}t}\left\vert \text{e}\right\rangle ,
\end{equation}%
after a long time.

Specifically, for an initial state $|\Phi (0)\rangle $ with the wave
function $\langle l|\Phi (0)\rangle $ varying slowly as $l$ changes, we have 
$c_{0}^{-}=0$\ and $|\Phi (\infty )\rangle \propto \left\vert \text{g}%
\right\rangle $. Then, the ground state $\left\vert \text{g}\right\rangle $\
can be dynamically generated from a variety of initial states. Note that the
analytical expressions of $\left\vert \text{g}\right\rangle $ and $%
\left\vert \text{e}\right\rangle $ from $|\psi _{0}^{\pm }\rangle $\ are
approximate. In practice, such two states can be obtained from numerical
simulations. To verify and demonstrate the above results, numerical
simulations of the dynamic process are performed using a uniform mesh for
the time discretization.

{We calculate the fidelity}

\begin{equation}
F(t)=\left\vert \left\langle \text{g}\right\vert e^{-iHt}|\Phi (0)\rangle
/\left\vert e^{-iHt}|\Phi (0)\rangle \right\vert \right\vert ^{2},
\label{Ft}
\end{equation}%
to measure the distance between the evolved state and the target state.{\ }%
We plot several typical initial states $\langle l|\Phi (0)\rangle $ and $F(t)
$\ in fig.\ \ref{fig3}. These numerical results are consistent with our
previous analysis, which showed that the evolved states converge to the
state $\left\vert \text{g}\right\rangle $ for several selected initial
states. {Therefore, we conclude that a harmonic imaginary potential can
support the stable bound state }$\left\vert \text{g}\right\rangle $, which
can also be prepared dynamically. {We also investigate the small
size systems by numerical simulation. We plot the profiles of ground states
and $F(t)$ of evolving from a uniform initial state in fig.\ \ref{fig4}. It
indicates that a small system size requires a small width of the
wavefunction profile, which requires a large $V$ (or $\omega$), resulting
in faster convergence. The small size and the faster convergence of the
fidelity make the scheme experimentally feasible.}

In general, the dynamics of a non-Hermitian two-level system are not Dirac
probability-preserving. However, as mentioned above, we have $\langle $e$|$g$%
\rangle =0$\ in the limit $\omega /J\rightarrow 0$ (or $\alpha \rightarrow 0$%
). This indicates that a superposition state obeys the Dirac
probability-preserving dynamics. We demonstrate this point by computing the
probability%
\begin{equation}
P(t)=\left\vert \langle \Phi (t)|\Phi (t)\rangle \right\vert ^{2}
\label{e26}
\end{equation}%
for initial state 
\begin{equation}
|\Phi (0)\rangle =\frac{1}{\sqrt{2}}\left( \left\vert \text{g}\right\rangle
+\left\vert \text{e}\right\rangle \right) ,
\end{equation}%
where $\left\vert \text{g}\right\rangle $\ and $\left\vert \text{e}%
\right\rangle $\ are obtained by the numerical simulation for the systems
with different values of $\omega /J$. The plots of the $P(t)$ function in
fig.\ \ref{fig5} conform to the theoretical predictions we established.

\section{Transition between two levels}

\label{Transition between two levels}

In parallel, one can dynamically generate the excited state $\left\vert 
\text{e}\right\rangle $\ from initial states with $c_{0}^{+}=0$. In this
section, we focus on a scheme designed to facilitate the transition between
the two states $\left\vert \text{g}\right\rangle $\ and $\left\vert \text{e}%
\right\rangle $. Based on the above analysis, a possible clue is provided by
the relation $\mathcal{PT}\left\vert \text{g}\right\rangle =\left\vert \text{%
e}\right\rangle $.\ We start with the dynamic realization of operator $%
\mathcal{P}$.

We consider to realize the action of operator $\mathcal{P}$ by the time
evolution under a quenching Hamiltonian%
\begin{equation}
H_{\mathrm{quen}}=H+H_{p}(t),  \label{e28}
\end{equation}%
where%
\begin{equation}
H_{p}(t)=\mu (t)\sum\limits_{l=-\infty }^{+\infty }l|l\rangle \langle l|,
\end{equation}%
describes the action of an extra time-dependent uniform field. Here the
coefficient is defined as%
\begin{equation}
\mu (t)=\left\{ 
\begin{array}{cc}
\frac{\pi }{\Delta }, & 0<t<\Delta \\ 
0, & \text{otherwise}%
\end{array}%
\right. ,
\end{equation}%
with which $H_{p}(t)$\ acts as a $\pi $\ pulse. In general, the time
evolution of a given initial state is governed by the time propagator%
\begin{equation}
U(t)=\exp \left[ -i\int_{0}^{t}H_{\mathrm{quen}}\mathrm{d}t\right] .
\end{equation}%
Under the condition $\pi /\Delta \gg 1$, we can have%
\begin{eqnarray}
U(t) &\approx &\exp \left[ -i\int_{0}^{t}H\mathrm{d}t\right] \exp \left[
-i\int_{0}^{\Delta }H_{p}(t)\mathrm{d}t\right]  \notag \\
&=&\exp \left[ -i\int_{0}^{t}H\mathrm{d}t\right] \prod\limits_{l=-\infty
}^{+\infty }\exp \left( -i\pi l\right) |l\rangle \langle l|  \notag \\
&=&\exp \left[ -i\int_{0}^{t}H\mathrm{d}t\right] \mathcal{P}.
\end{eqnarray}%
It indicates that the time evolution of the initial state $\left\vert \text{g%
}\right\rangle $ is%
\begin{equation}
U(t)\left\vert \text{g}\right\rangle =\exp \left[ -i\int_{0}^{t}H\mathrm{d}t%
\right] \left\vert \text{e}\right\rangle ^{\ast },
\end{equation}%
which tends to $\left\vert \text{e}\right\rangle $\ after a long time.
Inversely, we can also realize the operation%
\begin{equation}
U(t)\left\vert \text{e}\right\rangle \rightarrow \left\vert \text{g}%
\right\rangle .
\end{equation}

To verify this result, we perform numerical simulation for such a quenching
process for two different initial states $\left\vert \Phi (0)\right\rangle
=\left\vert \text{g}\right\rangle $ and $\left\vert \text{e}\right\rangle $,
which are obtained by exact diagonalization of the pre-quench Hamiltonian $H$%
. {We compute the two fidelities}

\begin{eqnarray}
F_{\text{g}}(t) &=&\left\vert \left\langle \text{g}\right\vert e^{-iHt}|%
\text{g}\rangle /\left\vert e^{-iHt}|\text{g}\rangle \right\vert \right\vert
^{2},  \label{Fg} \\
F_{\text{e}}(t) &=&\left\vert \left\langle \text{e}\right\vert e^{-iHt}|%
\text{g}\rangle /\left\vert e^{-iHt}|\text{g}\rangle \right\vert \right\vert
^{2},  \label{Fe}
\end{eqnarray}%
to evaluate the efficiency of the scheme. It is expected that the {fidelities%
} obey $F_{\text{g}}(\infty )=0$ and $F_{\text{e}}(\infty )=1$ (or $F_{\text{%
e}}(\infty )=0$ and $F_{\text{g}}(\infty )=1$). The numerical results are
presented in fig.\ \ref{fig6}(b), which are in accord with our analysis.
Here the form of $\mu (t)$\ is the simplest one for a $\pi $\ pulse. Other
types of $\mu (t)$\ could also have the same effect.

\section{Conclusion and Discussion}

\label{Conclusion and Discussion}

In summary, we have studied the formation of two-level bound states induced
by imaginary potential in a Hermitian tight-binding system. We have shown
that the real parts of energy levels are equidistant, while the imaginary
parts are semi-negative definite and equidistant. This allows for the
formation of an effective two-level system. The existence of bound states in
a system with an imaginary potential is fundamentally due to the fact that
the corresponding Hermitian system has a discrete spatial coordinates, which
arises from the presence of a periodic real potential. The similar
phenomenon occurs in another non-Hermitian system. It has been shown in Ref. 
\cite{Zhang2024} that bound states can form in a tight-binding chain in the
presence of a linear imaginary potential. This allows for the formation of
an effective two-level system. In contrast to a Hermitian two-level system,
a non-Hermitian system can have a single pair of real energy levels, while
the remaining levels are unstable due to their negative imaginary
components. This feature enables the dynamic preparation and operation of
two-level quantum states. Our findings reveal that the phenomenon of
dissipation gives rise to bound states that can effectively be described as
a two-level system.

\acknowledgments This work was supported by National Natural Science
Foundation of China (under Grant No. 12374461). 

\begin{thebibliography}{43}%
	\makeatletter
	\providecommand \@ifxundefined [1]{%
		\@ifx{#1\undefined}
	}%
	\providecommand \@ifnum [1]{%
		\ifnum #1\expandafter \@firstoftwo
		\else \expandafter \@secondoftwo
		\fi
	}%
	\providecommand \@ifx [1]{%
		\ifx #1\expandafter \@firstoftwo
		\else \expandafter \@secondoftwo
		\fi
	}%
	\providecommand \natexlab [1]{#1}%
	\providecommand \enquote  [1]{``#1''}%
	\providecommand \bibnamefont  [1]{#1}%
	\providecommand \bibfnamefont [1]{#1}%
	\providecommand \citenamefont [1]{#1}%
	\providecommand \href@noop [0]{\@secondoftwo}%
	\providecommand \href [0]{\begingroup \@sanitize@url \@href}%
	\providecommand \@href[1]{\@@startlink{#1}\@@href}%
	\providecommand \@@href[1]{\endgroup#1\@@endlink}%
	\providecommand \@sanitize@url [0]{\catcode `\\12\catcode `\$12\catcode
		`\&12\catcode `\#12\catcode `\^12\catcode `\_12\catcode `\%12\relax}%
	\providecommand \@@startlink[1]{}%
	\providecommand \@@endlink[0]{}%
	\providecommand \url  [0]{\begingroup\@sanitize@url \@url }%
	\providecommand \@url [1]{\endgroup\@href {#1}{\urlprefix }}%
	\providecommand \urlprefix  [0]{URL }%
	\providecommand \Eprint [0]{\href }%
	\providecommand \doibase [0]{http://dx.doi.org/}%
	\providecommand \selectlanguage [0]{\@gobble}%
	\providecommand \bibinfo  [0]{\@secondoftwo}%
	\providecommand \bibfield  [0]{\@secondoftwo}%
	\providecommand \translation [1]{[#1]}%
	\providecommand \BibitemOpen [0]{}%
	\providecommand \bibitemStop [0]{}%
	\providecommand \bibitemNoStop [0]{.\EOS\space}%
	\providecommand \EOS [0]{\spacefactor3000\relax}%
	\providecommand \BibitemShut  [1]{\csname bibitem#1\endcsname}%
	\let\auto@bib@innerbib\@empty
	\bibitem [{\citenamefont {Johnson}\ \emph {et~al.}(2001)\citenamefont
		{Johnson}, \citenamefont {Mekis}, \citenamefont {Fan},\ and\ \citenamefont
		{Joannopoulos}}]{Johnson2001}%
	\BibitemOpen
	\bibfield  {author} {\bibinfo {author} {\bibfnamefont {S.G.}\ \bibnamefont
			{Johnson}}, \bibinfo {author} {\bibfnamefont {A.}~\bibnamefont {Mekis}},
		\bibinfo {author} {\bibfnamefont {S.}~\bibnamefont {Fan}}, \ and\ \bibinfo
		{author} {\bibfnamefont {J.D.}\ \bibnamefont {Joannopoulos}},\ }\bibfield
	{title} {\enquote {\bibinfo {title} {Molding the flow of light},}\ }\href
	{\doibase 10.1109/5992.963426} {\bibfield  {journal} {\bibinfo  {journal}
			{Computing in Science amp; Engineering}\ }\textbf {\bibinfo {volume} {3}},\
		\bibinfo {pages} {38--47} (\bibinfo {year} {2001})}\BibitemShut {NoStop}%
	\bibitem [{\citenamefont {John}(1987)}]{John1987}%
	\BibitemOpen
	\bibfield  {author} {\bibinfo {author} {\bibfnamefont {Sajeev}\ \bibnamefont
			{John}},\ }\bibfield  {title} {\enquote {\bibinfo {title} {Strong
				localization of photons in certain disordered dielectric superlattices},}\
	}\href {\doibase 10.1103/physrevlett.58.2486} {\bibfield  {journal} {\bibinfo
			{journal} {Physical Review Letters}\ }\textbf {\bibinfo {volume} {58}},\
		\bibinfo {pages} {2486--2489} (\bibinfo {year} {1987})}\BibitemShut {NoStop}%
	\bibitem [{\citenamefont {Yablonovitch}(1987)}]{Yablonovitch1987}%
	\BibitemOpen
	\bibfield  {author} {\bibinfo {author} {\bibfnamefont {Eli}\ \bibnamefont
			{Yablonovitch}},\ }\bibfield  {title} {\enquote {\bibinfo {title} {{Inhibited
					Spontaneous Emission in Solid-State Physics and Electronics}},}\ }\href
	{\doibase 10.1103/physrevlett.58.2059} {\bibfield  {journal} {\bibinfo
			{journal} {Physical Review Letters}\ }\textbf {\bibinfo {volume} {58}},\
		\bibinfo {pages} {2059--2062} (\bibinfo {year} {1987})}\BibitemShut {NoStop}%
	\bibitem [{\citenamefont {Skorobogatiy}\ \emph {et~al.}(2005)\citenamefont
		{Skorobogatiy}, \citenamefont {Begin},\ and\ \citenamefont
		{Talneau}}]{Skorobogatiy2005}%
	\BibitemOpen
	\bibfield  {author} {\bibinfo {author} {\bibfnamefont {Maksim}\ \bibnamefont
			{Skorobogatiy}}, \bibinfo {author} {\bibfnamefont {Guillaume}\ \bibnamefont
			{Begin}}, \ and\ \bibinfo {author} {\bibfnamefont {Anne}\ \bibnamefont
			{Talneau}},\ }\bibfield  {title} {\enquote {\bibinfo {title} {Statistical
				analysis of geometrical imperfections from the images of 2d photonic
				crystals},}\ }\href {\doibase 10.1364/opex.13.002487} {\bibfield  {journal}
		{\bibinfo  {journal} {Optics Express}\ }\textbf {\bibinfo {volume} {13}},\
		\bibinfo {pages} {2487} (\bibinfo {year} {2005})}\BibitemShut {NoStop}%
	\bibitem [{\citenamefont {Engelen}\ \emph {et~al.}(2008)\citenamefont
		{Engelen}, \citenamefont {Mori}, \citenamefont {Baba},\ and\ \citenamefont
		{Kuipers}}]{Engelen2008}%
	\BibitemOpen
	\bibfield  {author} {\bibinfo {author} {\bibfnamefont {R.~J.~P.}\
			\bibnamefont {Engelen}}, \bibinfo {author} {\bibfnamefont {D.}~\bibnamefont
			{Mori}}, \bibinfo {author} {\bibfnamefont {T.}~\bibnamefont {Baba}}, \ and\
		\bibinfo {author} {\bibfnamefont {L.}~\bibnamefont {Kuipers}},\ }\bibfield
	{title} {\enquote {\bibinfo {title} {{Two Regimes of Slow-Light Losses
					Revealed by Adiabatic Reduction of Group Velocity}},}\ }\href {\doibase
		10.1103/physrevlett.101.103901} {\bibfield  {journal} {\bibinfo  {journal}
			{Physical Review Letters}\ }\textbf {\bibinfo {volume} {101}},\ \bibinfo
		{pages} {103901} (\bibinfo {year} {2008})}\BibitemShut {NoStop}%
	\bibitem [{\citenamefont {Hughes}\ \emph {et~al.}(2005)\citenamefont {Hughes},
		\citenamefont {Ramunno}, \citenamefont {Young},\ and\ \citenamefont
		{Sipe}}]{Hughes2005}%
	\BibitemOpen
	\bibfield  {author} {\bibinfo {author} {\bibfnamefont {S.}~\bibnamefont
			{Hughes}}, \bibinfo {author} {\bibfnamefont {L.}~\bibnamefont {Ramunno}},
		\bibinfo {author} {\bibfnamefont {Jeff~F.}\ \bibnamefont {Young}}, \ and\
		\bibinfo {author} {\bibfnamefont {J.~E.}\ \bibnamefont {Sipe}},\ }\bibfield
	{title} {\enquote {\bibinfo {title} {{Extrinsic Optical Scattering Loss in
					Photonic Crystal Waveguides: Role of Fabrication Disorder and Photon Group
					Velocity}},}\ }\href {\doibase 10.1103/physrevlett.94.033903} {\bibfield
		{journal} {\bibinfo  {journal} {Physical Review Letters}\ }\textbf {\bibinfo
			{volume} {94}},\ \bibinfo {pages} {033903} (\bibinfo {year}
		{2005})}\BibitemShut {NoStop}%
	\bibitem [{\citenamefont {Kuramochi}\ \emph {et~al.}(2005)\citenamefont
		{Kuramochi}, \citenamefont {Notomi}, \citenamefont {Hughes}, \citenamefont
		{Shinya}, \citenamefont {Watanabe},\ and\ \citenamefont
		{Ramunno}}]{Kuramochi2005}%
	\BibitemOpen
	\bibfield  {author} {\bibinfo {author} {\bibfnamefont {E.}~\bibnamefont
			{Kuramochi}}, \bibinfo {author} {\bibfnamefont {M.}~\bibnamefont {Notomi}},
		\bibinfo {author} {\bibfnamefont {S.}~\bibnamefont {Hughes}}, \bibinfo
		{author} {\bibfnamefont {A.}~\bibnamefont {Shinya}}, \bibinfo {author}
		{\bibfnamefont {T.}~\bibnamefont {Watanabe}}, \ and\ \bibinfo {author}
		{\bibfnamefont {L.}~\bibnamefont {Ramunno}},\ }\bibfield  {title} {\enquote
		{\bibinfo {title} {Disorder-induced scattering loss of line-defect waveguides
				in photonic crystal slabs},}\ }\href {\doibase 10.1103/physrevb.72.161318}
	{\bibfield  {journal} {\bibinfo  {journal} {Physical Review B}\ }\textbf
		{\bibinfo {volume} {72}},\ \bibinfo {pages} {161318} (\bibinfo {year}
		{2005})}\BibitemShut {NoStop}%
	\bibitem [{\citenamefont {Le~Thomas}\ \emph {et~al.}(2008)\citenamefont
		{Le~Thomas}, \citenamefont {Zabelin}, \citenamefont {Houdré}, \citenamefont
		{Kotlyar},\ and\ \citenamefont {Krauss}}]{LeThomas2008}%
	\BibitemOpen
	\bibfield  {author} {\bibinfo {author} {\bibfnamefont {N.}~\bibnamefont
			{Le~Thomas}}, \bibinfo {author} {\bibfnamefont {V.}~\bibnamefont {Zabelin}},
		\bibinfo {author} {\bibfnamefont {R.}~\bibnamefont {Houdré}}, \bibinfo
		{author} {\bibfnamefont {M.~V.}\ \bibnamefont {Kotlyar}}, \ and\ \bibinfo
		{author} {\bibfnamefont {T.~F.}\ \bibnamefont {Krauss}},\ }\bibfield  {title}
	{\enquote {\bibinfo {title} {{Influence of residual disorder on the
					anticrossing of Bloch modes probed in $\textbf{k}$ space}},}\ }\href
	{\doibase 10.1103/physrevb.78.125301} {\bibfield  {journal} {\bibinfo
			{journal} {Physical Review B}\ }\textbf {\bibinfo {volume} {78}},\ \bibinfo
		{pages} {125301} (\bibinfo {year} {2008})}\BibitemShut {NoStop}%
	\bibitem [{\citenamefont {Mazoyer}\ \emph {et~al.}(2009)\citenamefont
		{Mazoyer}, \citenamefont {Hugonin},\ and\ \citenamefont
		{Lalanne}}]{Mazoyer2009}%
	\BibitemOpen
	\bibfield  {author} {\bibinfo {author} {\bibfnamefont {S.}~\bibnamefont
			{Mazoyer}}, \bibinfo {author} {\bibfnamefont {J.~P.}\ \bibnamefont
			{Hugonin}}, \ and\ \bibinfo {author} {\bibfnamefont {P.}~\bibnamefont
			{Lalanne}},\ }\bibfield  {title} {\enquote {\bibinfo {title}
			{{Disorder-Induced Multiple Scattering in Photonic-Crystal Waveguides}},}\
	}\href {\doibase 10.1103/physrevlett.103.063903} {\bibfield  {journal}
		{\bibinfo  {journal} {Physical Review Letters}\ }\textbf {\bibinfo {volume}
			{103}},\ \bibinfo {pages} {063903} (\bibinfo {year} {2009})}\BibitemShut
	{NoStop}%
	\bibitem [{\citenamefont {Mazoyer}\ \emph {et~al.}(2010)\citenamefont
		{Mazoyer}, \citenamefont {Lalanne}, \citenamefont {Rodier}, \citenamefont
		{Hugonin}, \citenamefont {Spasenović}, \citenamefont {Kuipers},
		\citenamefont {Beggs},\ and\ \citenamefont {Krauss}}]{Mazoyer2010}%
	\BibitemOpen
	\bibfield  {author} {\bibinfo {author} {\bibfnamefont {S.}~\bibnamefont
			{Mazoyer}}, \bibinfo {author} {\bibfnamefont {P.}~\bibnamefont {Lalanne}},
		\bibinfo {author} {\bibfnamefont {J.C.}\ \bibnamefont {Rodier}}, \bibinfo
		{author} {\bibfnamefont {J.P.}\ \bibnamefont {Hugonin}}, \bibinfo {author}
		{\bibfnamefont {M.}~\bibnamefont {Spasenović}}, \bibinfo {author}
		{\bibfnamefont {L.}~\bibnamefont {Kuipers}}, \bibinfo {author} {\bibfnamefont
			{D.M.}\ \bibnamefont {Beggs}}, \ and\ \bibinfo {author} {\bibfnamefont
			{T.F.}\ \bibnamefont {Krauss}},\ }\bibfield  {title} {\enquote {\bibinfo
			{title} {Statistical fluctuations of transmission in slow light
				photonic-crystal waveguides},}\ }\href {\doibase 10.1364/oe.18.014654}
	{\bibfield  {journal} {\bibinfo  {journal} {Optics Express}\ }\textbf
		{\bibinfo {volume} {18}},\ \bibinfo {pages} {14654} (\bibinfo {year}
		{2010})}\BibitemShut {NoStop}%
	\bibitem [{\citenamefont {O’Faolain}\ \emph {et~al.}(2007)\citenamefont
		{O’Faolain}, \citenamefont {White}, \citenamefont {O’Brien},
		\citenamefont {Yuan}, \citenamefont {Settle},\ and\ \citenamefont
		{Krauss}}]{O’Faolain2007}%
	\BibitemOpen
	\bibfield  {author} {\bibinfo {author} {\bibfnamefont {Liam}\ \bibnamefont
			{O’Faolain}}, \bibinfo {author} {\bibfnamefont {Thomas~P.}\ \bibnamefont
			{White}}, \bibinfo {author} {\bibfnamefont {David}\ \bibnamefont
			{O’Brien}}, \bibinfo {author} {\bibfnamefont {Xiadong}\ \bibnamefont
			{Yuan}}, \bibinfo {author} {\bibfnamefont {Michael~D.}\ \bibnamefont
			{Settle}}, \ and\ \bibinfo {author} {\bibfnamefont {Thomas~F.}\ \bibnamefont
			{Krauss}},\ }\bibfield  {title} {\enquote {\bibinfo {title} {Dependence of
				extrinsic loss on group velocity in photonic crystal waveguides},}\ }\href
	{\doibase 10.1364/oe.15.013129} {\bibfield  {journal} {\bibinfo  {journal}
			{Optics Express}\ }\textbf {\bibinfo {volume} {15}},\ \bibinfo {pages}
		{13129} (\bibinfo {year} {2007})}\BibitemShut {NoStop}%
	\bibitem [{\citenamefont {Bender}\ and\ \citenamefont
		{Boettcher}(1998)}]{Bender1998}%
	\BibitemOpen
	\bibfield  {author} {\bibinfo {author} {\bibfnamefont {Carl~M.}\ \bibnamefont
			{Bender}}\ and\ \bibinfo {author} {\bibfnamefont {Stefan}\ \bibnamefont
			{Boettcher}},\ }\bibfield  {title} {\enquote {\bibinfo {title} {{Real Spectra
					in Non-Hermitian Hamiltonians Having $\mathcal{PT}$ Symmetry}},}\ }\href
	{\doibase 10.1103/physrevlett.80.5243} {\bibfield  {journal} {\bibinfo
			{journal} {Physical Review Letters}\ }\textbf {\bibinfo {volume} {80}},\
		\bibinfo {pages} {5243--5246} (\bibinfo {year} {1998})}\BibitemShut {NoStop}%
	\bibitem [{\citenamefont {Dorey}\ \emph {et~al.}(2001)\citenamefont {Dorey},
		\citenamefont {Dunning},\ and\ \citenamefont {Tateo}}]{Dorey2001}%
	\BibitemOpen
	\bibfield  {author} {\bibinfo {author} {\bibfnamefont {Patrick}\ \bibnamefont
			{Dorey}}, \bibinfo {author} {\bibfnamefont {Clare}\ \bibnamefont {Dunning}},
		\ and\ \bibinfo {author} {\bibfnamefont {Roberto}\ \bibnamefont {Tateo}},\
	}\bibfield  {title} {\enquote {\bibinfo {title} {Spectral equivalences, bethe
				ansatz equations, and reality properties in $\mathcal{PT}$-symmetric quantum
				mechanics},}\ }\href {\doibase 10.1088/0305-4470/34/28/305} {\bibfield
		{journal} {\bibinfo  {journal} {Journal of Physics A: Mathematical and
				General}\ }\textbf {\bibinfo {volume} {34}},\ \bibinfo {pages} {5679--5704}
		(\bibinfo {year} {2001})}\BibitemShut {NoStop}%
	\bibitem [{\citenamefont {Mostafazadeh}(2002)}]{Mostafazadeh2002}%
	\BibitemOpen
	\bibfield  {author} {\bibinfo {author} {\bibfnamefont {Ali}\ \bibnamefont
			{Mostafazadeh}},\ }\bibfield  {title} {\enquote {\bibinfo {title}
			{{Pseudo-Hermiticity versus PT symmetry: the necessary condition for the
					reality of the spectrum of a non-Hermitian Hamiltonian}},}\ }\href@noop {}
	{\bibfield  {journal} {\bibinfo  {journal} {Journal of Mathematical Physics}\
		}\textbf {\bibinfo {volume} {43}},\ \bibinfo {pages} {205--214} (\bibinfo
		{year} {2002})}\BibitemShut {NoStop}%
	\bibitem [{\citenamefont {Znojil}(1999)}]{Znojil1999}%
	\BibitemOpen
	\bibfield  {author} {\bibinfo {author} {\bibfnamefont {Miloslav}\
			\bibnamefont {Znojil}},\ }\bibfield  {title} {\enquote {\bibinfo {title}
			{{PT-symmetric harmonic oscillators}},}\ }\href {\doibase
		10.1016/s0375-9601(99)00429-6} {\bibfield  {journal} {\bibinfo  {journal}
			{Physics Letters A}\ }\textbf {\bibinfo {volume} {259}},\ \bibinfo {pages}
		{220--223} (\bibinfo {year} {1999})}\BibitemShut {NoStop}%
	\bibitem [{\citenamefont {Jones}(2005)}]{Jones2005}%
	\BibitemOpen
	\bibfield  {author} {\bibinfo {author} {\bibfnamefont {H~F}\ \bibnamefont
			{Jones}},\ }\bibfield  {title} {\enquote {\bibinfo {title} {{On
					pseudo-Hermitian Hamiltonians and their Hermitian counterparts}},}\ }\href
	{\doibase 10.1088/0305-4470/38/8/010} {\bibfield  {journal} {\bibinfo
			{journal} {Journal of Physics A: Mathematical and General}\ }\textbf
		{\bibinfo {volume} {38}},\ \bibinfo {pages} {1741--1746} (\bibinfo {year}
		{2005})}\BibitemShut {NoStop}%
	\bibitem [{\citenamefont {El-Ganainy}\ \emph {et~al.}(2007)\citenamefont
		{El-Ganainy}, \citenamefont {Makris}, \citenamefont {Christodoulides},\ and\
		\citenamefont {Musslimani}}]{ElGanainy2007}%
	\BibitemOpen
	\bibfield  {author} {\bibinfo {author} {\bibfnamefont {R.}~\bibnamefont
			{El-Ganainy}}, \bibinfo {author} {\bibfnamefont {K.~G.}\ \bibnamefont
			{Makris}}, \bibinfo {author} {\bibfnamefont {D.~N.}\ \bibnamefont
			{Christodoulides}}, \ and\ \bibinfo {author} {\bibfnamefont {Ziad~H.}\
			\bibnamefont {Musslimani}},\ }\bibfield  {title} {\enquote {\bibinfo {title}
			{{Theory of coupled optical PT-symmetric structures}},}\ }\href {\doibase
		10.1364/ol.32.002632} {\bibfield  {journal} {\bibinfo  {journal} {Optics
				Letters}\ }\textbf {\bibinfo {volume} {32}},\ \bibinfo {pages} {2632}
		(\bibinfo {year} {2007})}\BibitemShut {NoStop}%
	\bibitem [{\citenamefont {Musslimani}\ \emph {et~al.}(2008)\citenamefont
		{Musslimani}, \citenamefont {Makris}, \citenamefont {El-Ganainy},\ and\
		\citenamefont {Christodoulides}}]{Musslimani2008}%
	\BibitemOpen
	\bibfield  {author} {\bibinfo {author} {\bibfnamefont {Z.~H.}\ \bibnamefont
			{Musslimani}}, \bibinfo {author} {\bibfnamefont {K.~G.}\ \bibnamefont
			{Makris}}, \bibinfo {author} {\bibfnamefont {R.}~\bibnamefont {El-Ganainy}},
		\ and\ \bibinfo {author} {\bibfnamefont {D.~N.}\ \bibnamefont
			{Christodoulides}},\ }\bibfield  {title} {\enquote {\bibinfo {title}
			{{Optical Solitons in $\mathcal{PT}$ Periodic Potentials}},}\ }\href
	{\doibase 10.1103/physrevlett.100.030402} {\bibfield  {journal} {\bibinfo
			{journal} {Physical Review Letters}\ }\textbf {\bibinfo {volume} {100}},\
		\bibinfo {pages} {030402} (\bibinfo {year} {2008})}\BibitemShut {NoStop}%
	\bibitem [{\citenamefont {Makris}\ \emph {et~al.}(2008)\citenamefont {Makris},
		\citenamefont {El-Ganainy}, \citenamefont {Christodoulides},\ and\
		\citenamefont {Musslimani}}]{Makris2008}%
	\BibitemOpen
	\bibfield  {author} {\bibinfo {author} {\bibfnamefont {K.~G.}\ \bibnamefont
			{Makris}}, \bibinfo {author} {\bibfnamefont {R.}~\bibnamefont {El-Ganainy}},
		\bibinfo {author} {\bibfnamefont {D.~N.}\ \bibnamefont {Christodoulides}}, \
		and\ \bibinfo {author} {\bibfnamefont {Z.~H.}\ \bibnamefont {Musslimani}},\
	}\bibfield  {title} {\enquote {\bibinfo {title} {{Beam Dynamics in
					$\mathcal{PT}$ Symmetric Optical Lattices}},}\ }\href {\doibase
		10.1103/physrevlett.100.103904} {\bibfield  {journal} {\bibinfo  {journal}
			{Physical Review Letters}\ }\textbf {\bibinfo {volume} {100}},\ \bibinfo
		{pages} {103904} (\bibinfo {year} {2008})}\BibitemShut {NoStop}%
	\bibitem [{\citenamefont {Joglekar}\ \emph {et~al.}(2010)\citenamefont
		{Joglekar}, \citenamefont {Scott}, \citenamefont {Babbey},\ and\
		\citenamefont {Saxena}}]{Joglekar2010}%
	\BibitemOpen
	\bibfield  {author} {\bibinfo {author} {\bibfnamefont {Yogesh~N.}\
			\bibnamefont {Joglekar}}, \bibinfo {author} {\bibfnamefont {Derek}\
			\bibnamefont {Scott}}, \bibinfo {author} {\bibfnamefont {Mark}\ \bibnamefont
			{Babbey}}, \ and\ \bibinfo {author} {\bibfnamefont {Avadh}\ \bibnamefont
			{Saxena}},\ }\bibfield  {title} {\enquote {\bibinfo {title} {Robust and
				fragile $\mathcal{PT}$-symmetric phases in a tight-binding chain},}\ }\href
	{\doibase 10.1103/physreva.82.030103} {\bibfield  {journal} {\bibinfo
			{journal} {Physical Review A}\ }\textbf {\bibinfo {volume} {82}},\ \bibinfo
		{pages} {030103} (\bibinfo {year} {2010})}\BibitemShut {NoStop}%
	\bibitem [{\citenamefont {Scott}\ and\ \citenamefont
		{Joglekar}(2011)}]{Scott2011}%
	\BibitemOpen
	\bibfield  {author} {\bibinfo {author} {\bibfnamefont {Derek~D.}\
			\bibnamefont {Scott}}\ and\ \bibinfo {author} {\bibfnamefont {Yogesh~N.}\
			\bibnamefont {Joglekar}},\ }\bibfield  {title} {\enquote {\bibinfo {title}
			{Degrees and signatures of broken $\mathcal{PT}$ symmetry in nonuniform
				lattices},}\ }\href {\doibase 10.1103/physreva.83.050102} {\bibfield
		{journal} {\bibinfo  {journal} {Physical Review A}\ }\textbf {\bibinfo
			{volume} {83}},\ \bibinfo {pages} {050102} (\bibinfo {year}
		{2011})}\BibitemShut {NoStop}%
	\bibitem [{\citenamefont {Chong}\ \emph {et~al.}(2010)\citenamefont {Chong},
		\citenamefont {Ge}, \citenamefont {Cao},\ and\ \citenamefont
		{Stone}}]{Chong2010}%
	\BibitemOpen
	\bibfield  {author} {\bibinfo {author} {\bibfnamefont {Y.~D.}\ \bibnamefont
			{Chong}}, \bibinfo {author} {\bibfnamefont {Li}~\bibnamefont {Ge}}, \bibinfo
		{author} {\bibfnamefont {Hui}\ \bibnamefont {Cao}}, \ and\ \bibinfo {author}
		{\bibfnamefont {A.~D.}\ \bibnamefont {Stone}},\ }\bibfield  {title} {\enquote
		{\bibinfo {title} {{Coherent Perfect Absorbers: Time-Reversed Lasers}},}\
	}\href {\doibase 10.1103/physrevlett.105.053901} {\bibfield  {journal}
		{\bibinfo  {journal} {Physical Review Letters}\ }\textbf {\bibinfo {volume}
			{105}},\ \bibinfo {pages} {053901} (\bibinfo {year} {2010})}\BibitemShut
	{NoStop}%
	\bibitem [{\citenamefont {Jing}\ \emph {et~al.}(2014)\citenamefont {Jing},
		\citenamefont {Özdemir}, \citenamefont {Lü}, \citenamefont {Zhang},
		\citenamefont {Yang},\ and\ \citenamefont {Nori}}]{Jing2014}%
	\BibitemOpen
	\bibfield  {author} {\bibinfo {author} {\bibfnamefont {Hui}\ \bibnamefont
			{Jing}}, \bibinfo {author} {\bibfnamefont {S.K.}\ \bibnamefont
			{Özdemir}}, \bibinfo {author} {\bibfnamefont {Xin-You}\ \bibnamefont {Lü}},
		\bibinfo {author} {\bibfnamefont {Jing}\ \bibnamefont {Zhang}}, \bibinfo
		{author} {\bibfnamefont {Lan}\ \bibnamefont {Yang}}, \ and\ \bibinfo {author}
		{\bibfnamefont {Franco}\ \bibnamefont {Nori}},\ }\bibfield  {title} {\enquote
		{\bibinfo {title} {{$\mathcal{PT}$-Symmetric Phonon Laser}},}\ }\href
	{\doibase 10.1103/physrevlett.113.053604} {\bibfield  {journal} {\bibinfo
			{journal} {Physical Review Letters}\ }\textbf {\bibinfo {volume} {113}},\
		\bibinfo {pages} {053604} (\bibinfo {year} {2014})}\BibitemShut {NoStop}%
	\bibitem [{\citenamefont {Guo}\ \emph {et~al.}(2009)\citenamefont {Guo},
		\citenamefont {Salamo}, \citenamefont {Duchesne}, \citenamefont {Morandotti},
		\citenamefont {Volatier-Ravat}, \citenamefont {Aimez}, \citenamefont
		{Siviloglou},\ and\ \citenamefont {Christodoulides}}]{Guo2009}%
	\BibitemOpen
	\bibfield  {author} {\bibinfo {author} {\bibfnamefont {A.}~\bibnamefont
			{Guo}}, \bibinfo {author} {\bibfnamefont {G.~J.}\ \bibnamefont {Salamo}},
		\bibinfo {author} {\bibfnamefont {D.}~\bibnamefont {Duchesne}}, \bibinfo
		{author} {\bibfnamefont {R.}~\bibnamefont {Morandotti}}, \bibinfo {author}
		{\bibfnamefont {M.}~\bibnamefont {Volatier-Ravat}}, \bibinfo {author}
		{\bibfnamefont {V.}~\bibnamefont {Aimez}}, \bibinfo {author} {\bibfnamefont
			{G.~A.}\ \bibnamefont {Siviloglou}}, \ and\ \bibinfo {author} {\bibfnamefont
			{D.~N.}\ \bibnamefont {Christodoulides}},\ }\bibfield  {title} {\enquote
		{\bibinfo {title} {{Observation of $\mathcal{PT}$-Symmetry Breaking in
					Complex Optical Potentials}},}\ }\href {\doibase
		10.1103/physrevlett.103.093902} {\bibfield  {journal} {\bibinfo  {journal}
			{Physical Review Letters}\ }\textbf {\bibinfo {volume} {103}},\ \bibinfo
		{pages} {093902} (\bibinfo {year} {2009})}\BibitemShut {NoStop}%
	\bibitem [{\citenamefont {Rüter}\ \emph {et~al.}(2010)\citenamefont {Rüter},
		\citenamefont {Makris}, \citenamefont {El-Ganainy}, \citenamefont
		{Christodoulides}, \citenamefont {Segev},\ and\ \citenamefont
		{Kip}}]{Rueter2010}%
	\BibitemOpen
	\bibfield  {author} {\bibinfo {author} {\bibfnamefont {Christian~E.}\
			\bibnamefont {Rüter}}, \bibinfo {author} {\bibfnamefont {Konstantinos~G.}\
			\bibnamefont {Makris}}, \bibinfo {author} {\bibfnamefont {Ramy}\ \bibnamefont
			{El-Ganainy}}, \bibinfo {author} {\bibfnamefont {Demetrios~N.}\ \bibnamefont
			{Christodoulides}}, \bibinfo {author} {\bibfnamefont {Mordechai}\
			\bibnamefont {Segev}}, \ and\ \bibinfo {author} {\bibfnamefont {Detlef}\
			\bibnamefont {Kip}},\ }\bibfield  {title} {\enquote {\bibinfo {title}
			{Observation of parity–time symmetry in optics},}\ }\href {\doibase
		10.1038/nphys1515} {\bibfield  {journal} {\bibinfo  {journal} {Nature
				Physics}\ }\textbf {\bibinfo {volume} {6}},\ \bibinfo {pages} {192--195}
		(\bibinfo {year} {2010})}\BibitemShut {NoStop}%
	\bibitem [{\citenamefont {Wan}\ \emph {et~al.}(2011)\citenamefont {Wan},
		\citenamefont {Chong}, \citenamefont {Ge}, \citenamefont {Noh}, \citenamefont
		{Stone},\ and\ \citenamefont {Cao}}]{Wan2011}%
	\BibitemOpen
	\bibfield  {author} {\bibinfo {author} {\bibfnamefont {Wenjie}\ \bibnamefont
			{Wan}}, \bibinfo {author} {\bibfnamefont {Yidong}\ \bibnamefont {Chong}},
		\bibinfo {author} {\bibfnamefont {Li}~\bibnamefont {Ge}}, \bibinfo {author}
		{\bibfnamefont {Heeso}\ \bibnamefont {Noh}}, \bibinfo {author} {\bibfnamefont
			{A.~Douglas}\ \bibnamefont {Stone}}, \ and\ \bibinfo {author} {\bibfnamefont
			{Hui}\ \bibnamefont {Cao}},\ }\bibfield  {title} {\enquote {\bibinfo {title}
			{{Time-Reversed Lasing and Interferometric Control of Absorption}},}\ }\href
	{\doibase 10.1126/science.1200735} {\bibfield  {journal} {\bibinfo  {journal}
			{Science}\ }\textbf {\bibinfo {volume} {331}},\ \bibinfo {pages} {889--892}
		(\bibinfo {year} {2011})}\BibitemShut {NoStop}%
	\bibitem [{\citenamefont {Sun}\ \emph {et~al.}(2014)\citenamefont {Sun},
		\citenamefont {Tan}, \citenamefont {Li}, \citenamefont {Li},\ and\
		\citenamefont {Chen}}]{Sun2014}%
	\BibitemOpen
	\bibfield  {author} {\bibinfo {author} {\bibfnamefont {Yong}\ \bibnamefont
			{Sun}}, \bibinfo {author} {\bibfnamefont {Wei}\ \bibnamefont {Tan}}, \bibinfo
		{author} {\bibfnamefont {Hong-qiang}\ \bibnamefont {Li}}, \bibinfo {author}
		{\bibfnamefont {Jensen}\ \bibnamefont {Li}}, \ and\ \bibinfo {author}
		{\bibfnamefont {Hong}\ \bibnamefont {Chen}},\ }\bibfield  {title} {\enquote
		{\bibinfo {title} {{Experimental Demonstration of a Coherent Perfect Absorber
					with PT Phase Transition}},}\ }\href {\doibase
		10.1103/physrevlett.112.143903} {\bibfield  {journal} {\bibinfo  {journal}
			{Physical Review Letters}\ }\textbf {\bibinfo {volume} {112}},\ \bibinfo
		{pages} {143903} (\bibinfo {year} {2014})}\BibitemShut {NoStop}%
	\bibitem [{\citenamefont {Feng}\ \emph {et~al.}(2012)\citenamefont {Feng},
		\citenamefont {Xu}, \citenamefont {Fegadolli}, \citenamefont {Lu},
		\citenamefont {Oliveira}, \citenamefont {Almeida}, \citenamefont {Chen},\
		and\ \citenamefont {Scherer}}]{Feng2012}%
	\BibitemOpen
	\bibfield  {author} {\bibinfo {author} {\bibfnamefont {Liang}\ \bibnamefont
			{Feng}}, \bibinfo {author} {\bibfnamefont {Ye-Long}\ \bibnamefont {Xu}},
		\bibinfo {author} {\bibfnamefont {William~S.}\ \bibnamefont {Fegadolli}},
		\bibinfo {author} {\bibfnamefont {Ming-Hui}\ \bibnamefont {Lu}}, \bibinfo
		{author} {\bibfnamefont {José E.~B.}\ \bibnamefont {Oliveira}}, \bibinfo
		{author} {\bibfnamefont {Vilson~R.}\ \bibnamefont {Almeida}}, \bibinfo
		{author} {\bibfnamefont {Yan-Feng}\ \bibnamefont {Chen}}, \ and\ \bibinfo
		{author} {\bibfnamefont {Axel}\ \bibnamefont {Scherer}},\ }\bibfield  {title}
	{\enquote {\bibinfo {title} {Experimental demonstration of a unidirectional
				reflectionless parity-time metamaterial at optical frequencies},}\ }\href
	{\doibase 10.1038/nmat3495} {\bibfield  {journal} {\bibinfo  {journal}
			{Nature Materials}\ }\textbf {\bibinfo {volume} {12}},\ \bibinfo {pages}
		{108--113} (\bibinfo {year} {2012})}\BibitemShut {NoStop}%
	\bibitem [{\citenamefont {Peng}\ \emph {et~al.}(2014)\citenamefont {Peng},
		\citenamefont {Özdemir}, \citenamefont {Lei}, \citenamefont {Monifi},
		\citenamefont {Gianfreda}, \citenamefont {Long}, \citenamefont {Fan},
		\citenamefont {Nori}, \citenamefont {Bender},\ and\ \citenamefont
		{Yang}}]{Peng2014}%
	\BibitemOpen
	\bibfield  {author} {\bibinfo {author} {\bibfnamefont {Bo}~\bibnamefont
			{Peng}}, \bibinfo {author} {\bibfnamefont {Şahin~Kaya}\ \bibnamefont
			{Özdemir}}, \bibinfo {author} {\bibfnamefont {Fuchuan}\ \bibnamefont {Lei}},
		\bibinfo {author} {\bibfnamefont {Faraz}\ \bibnamefont {Monifi}}, \bibinfo
		{author} {\bibfnamefont {Mariagiovanna}\ \bibnamefont {Gianfreda}}, \bibinfo
		{author} {\bibfnamefont {Gui~Lu}\ \bibnamefont {Long}}, \bibinfo {author}
		{\bibfnamefont {Shanhui}\ \bibnamefont {Fan}}, \bibinfo {author}
		{\bibfnamefont {Franco}\ \bibnamefont {Nori}}, \bibinfo {author}
		{\bibfnamefont {Carl~M.}\ \bibnamefont {Bender}}, \ and\ \bibinfo {author}
		{\bibfnamefont {Lan}\ \bibnamefont {Yang}},\ }\bibfield  {title} {\enquote
		{\bibinfo {title} {Parity–time-symmetric whispering-gallery
				microcavities},}\ }\href {\doibase 10.1038/nphys2927} {\bibfield  {journal}
		{\bibinfo  {journal} {Nature Physics}\ }\textbf {\bibinfo {volume} {10}},\
		\bibinfo {pages} {394--398} (\bibinfo {year} {2014})}\BibitemShut {NoStop}%
	\bibitem [{\citenamefont {Chang}\ \emph {et~al.}(2014)\citenamefont {Chang},
		\citenamefont {Jiang}, \citenamefont {Hua}, \citenamefont {Yang},
		\citenamefont {Wen}, \citenamefont {Jiang}, \citenamefont {Li}, \citenamefont
		{Wang},\ and\ \citenamefont {Xiao}}]{Chang2014}%
	\BibitemOpen
	\bibfield  {author} {\bibinfo {author} {\bibfnamefont {Long}\ \bibnamefont
			{Chang}}, \bibinfo {author} {\bibfnamefont {Xiaoshun}\ \bibnamefont {Jiang}},
		\bibinfo {author} {\bibfnamefont {Shiyue}\ \bibnamefont {Hua}}, \bibinfo
		{author} {\bibfnamefont {Chao}\ \bibnamefont {Yang}}, \bibinfo {author}
		{\bibfnamefont {Jianming}\ \bibnamefont {Wen}}, \bibinfo {author}
		{\bibfnamefont {Liang}\ \bibnamefont {Jiang}}, \bibinfo {author}
		{\bibfnamefont {Guanyu}\ \bibnamefont {Li}}, \bibinfo {author} {\bibfnamefont
			{Guanzhong}\ \bibnamefont {Wang}}, \ and\ \bibinfo {author} {\bibfnamefont
			{Min}\ \bibnamefont {Xiao}},\ }\bibfield  {title} {\enquote {\bibinfo {title}
			{Parity–time symmetry and variable optical isolation in
				active–passive-coupled microresonators},}\ }\href {\doibase
		10.1038/nphoton.2014.133} {\bibfield  {journal} {\bibinfo  {journal} {Nature
				Photonics}\ }\textbf {\bibinfo {volume} {8}},\ \bibinfo {pages} {524--529}
		(\bibinfo {year} {2014})}\BibitemShut {NoStop}%
	\bibitem [{\citenamefont {Feng}\ \emph {et~al.}(2014)\citenamefont {Feng},
		\citenamefont {Wong}, \citenamefont {Ma}, \citenamefont {Wang},\ and\
		\citenamefont {Zhang}}]{Feng2014}%
	\BibitemOpen
	\bibfield  {author} {\bibinfo {author} {\bibfnamefont {Liang}\ \bibnamefont
			{Feng}}, \bibinfo {author} {\bibfnamefont {Zi~Jing}\ \bibnamefont {Wong}},
		\bibinfo {author} {\bibfnamefont {Ren-Min}\ \bibnamefont {Ma}}, \bibinfo
		{author} {\bibfnamefont {Yuan}\ \bibnamefont {Wang}}, \ and\ \bibinfo
		{author} {\bibfnamefont {Xiang}\ \bibnamefont {Zhang}},\ }\bibfield  {title}
	{\enquote {\bibinfo {title} {Single-mode laser by parity-time symmetry
				breaking},}\ }\href {\doibase 10.1126/science.1258479} {\bibfield  {journal}
		{\bibinfo  {journal} {Science}\ }\textbf {\bibinfo {volume} {346}},\ \bibinfo
		{pages} {972--975} (\bibinfo {year} {2014})}\BibitemShut {NoStop}%
	\bibitem [{\citenamefont {Hodaei}\ \emph {et~al.}(2014)\citenamefont {Hodaei},
		\citenamefont {Miri}, \citenamefont {Heinrich}, \citenamefont
		{Christodoulides},\ and\ \citenamefont {Khajavikhan}}]{Hodaei2014}%
	\BibitemOpen
	\bibfield  {author} {\bibinfo {author} {\bibfnamefont {Hossein}\ \bibnamefont
			{Hodaei}}, \bibinfo {author} {\bibfnamefont {Mohammad-Ali}\ \bibnamefont
			{Miri}}, \bibinfo {author} {\bibfnamefont {Matthias}\ \bibnamefont
			{Heinrich}}, \bibinfo {author} {\bibfnamefont {Demetrios~N.}\ \bibnamefont
			{Christodoulides}}, \ and\ \bibinfo {author} {\bibfnamefont {Mercedeh}\
			\bibnamefont {Khajavikhan}},\ }\bibfield  {title} {\enquote {\bibinfo {title}
			{Parity-time–symmetric microring lasers},}\ }\href {\doibase
		10.1126/science.1258480} {\bibfield  {journal} {\bibinfo  {journal}
			{Science}\ }\textbf {\bibinfo {volume} {346}},\ \bibinfo {pages} {975--978}
		(\bibinfo {year} {2014})}\BibitemShut {NoStop}%
	\bibitem [{\citenamefont {Wimmer}\ \emph {et~al.}(2015)\citenamefont {Wimmer},
		\citenamefont {Regensburger}, \citenamefont {Miri}, \citenamefont {Bersch},
		\citenamefont {Christodoulides},\ and\ \citenamefont {Peschel}}]{Wimmer2015}%
	\BibitemOpen
	\bibfield  {author} {\bibinfo {author} {\bibfnamefont {Martin}\ \bibnamefont
			{Wimmer}}, \bibinfo {author} {\bibfnamefont {Alois}\ \bibnamefont
			{Regensburger}}, \bibinfo {author} {\bibfnamefont {Mohammad-Ali}\
			\bibnamefont {Miri}}, \bibinfo {author} {\bibfnamefont {Christoph}\
			\bibnamefont {Bersch}}, \bibinfo {author} {\bibfnamefont {Demetrios~N.}\
			\bibnamefont {Christodoulides}}, \ and\ \bibinfo {author} {\bibfnamefont
			{Ulf}\ \bibnamefont {Peschel}},\ }\bibfield  {title} {\enquote {\bibinfo
			{title} {Observation of optical solitons in pt-symmetric lattices},}\ }\href
	{\doibase 10.1038/ncomms8782} {\bibfield  {journal} {\bibinfo  {journal}
			{Nature Communications}\ }\textbf {\bibinfo {volume} {6}} (\bibinfo {year}
		{2015}),\ 10.1038/ncomms8782}\BibitemShut {NoStop}%
	\bibitem [{\citenamefont {Zhang}\ and\ \citenamefont {Song}(2013)}]{Zhang2013}%
	\BibitemOpen
	\bibfield  {author} {\bibinfo {author} {\bibfnamefont {X.Z.}\ \bibnamefont
			{Zhang}}\ and\ \bibinfo {author} {\bibfnamefont {Z.}~\bibnamefont {Song}},\
	}\bibfield  {title} {\enquote {\bibinfo {title} {{Momentum-independent
					reflectionless transmission in the non-Hermitian time-reversal symmetric
					system}},}\ }\href {\doibase 10.1016/j.aop.2013.08.012} {\bibfield  {journal}
		{\bibinfo  {journal} {Annals of Physics}\ }\textbf {\bibinfo {volume}
			{339}},\ \bibinfo {pages} {109--121} (\bibinfo {year} {2013})}\BibitemShut
	{NoStop}%
	\bibitem [{\citenamefont {Kunst}\ \emph {et~al.}(2018)\citenamefont {Kunst},
		\citenamefont {Edvardsson}, \citenamefont {Budich},\ and\ \citenamefont
		{Bergholtz}}]{Kunst2018}%
	\BibitemOpen
	\bibfield  {author} {\bibinfo {author} {\bibfnamefont {Flore~K.}\
			\bibnamefont {Kunst}}, \bibinfo {author} {\bibfnamefont {Elisabet}\
			\bibnamefont {Edvardsson}}, \bibinfo {author} {\bibfnamefont {Jan~Carl}\
			\bibnamefont {Budich}}, \ and\ \bibinfo {author} {\bibfnamefont {Emil~J.}\
			\bibnamefont {Bergholtz}},\ }\bibfield  {title} {\enquote {\bibinfo {title}
			{{Biorthogonal Bulk-Boundary Correspondence in Non-Hermitian Systems}},}\
	}\href {\doibase 10.1103/physrevlett.121.026808} {\bibfield  {journal}
		{\bibinfo  {journal} {Physical Review Letters}\ }\textbf {\bibinfo {volume}
			{121}},\ \bibinfo {pages} {026808} (\bibinfo {year} {2018})}\BibitemShut
	{NoStop}%
	\bibitem [{\citenamefont {Yao}\ and\ \citenamefont {Wang}(2018)}]{Yao2018}%
	\BibitemOpen
	\bibfield  {author} {\bibinfo {author} {\bibfnamefont {Shunyu}\ \bibnamefont
			{Yao}}\ and\ \bibinfo {author} {\bibfnamefont {Zhong}\ \bibnamefont {Wang}},\
	}\bibfield  {title} {\enquote {\bibinfo {title} {{Edge States and Topological
					Invariants of Non-Hermitian Systems}},}\ }\href {\doibase
		10.1103/physrevlett.121.086803} {\bibfield  {journal} {\bibinfo  {journal}
			{Physical Review Letters}\ }\textbf {\bibinfo {volume} {121}},\ \bibinfo
		{pages} {086803} (\bibinfo {year} {2018})}\BibitemShut {NoStop}%
	\bibitem [{\citenamefont {Gong}\ \emph {et~al.}(2018)\citenamefont {Gong},
		\citenamefont {Ashida}, \citenamefont {Kawabata}, \citenamefont {Takasan},
		\citenamefont {Higashikawa},\ and\ \citenamefont {Ueda}}]{Gong2018}%
	\BibitemOpen
	\bibfield  {author} {\bibinfo {author} {\bibfnamefont {Zongping}\
			\bibnamefont {Gong}}, \bibinfo {author} {\bibfnamefont {Yuto}\ \bibnamefont
			{Ashida}}, \bibinfo {author} {\bibfnamefont {Kohei}\ \bibnamefont
			{Kawabata}}, \bibinfo {author} {\bibfnamefont {Kazuaki}\ \bibnamefont
			{Takasan}}, \bibinfo {author} {\bibfnamefont {Sho}\ \bibnamefont
			{Higashikawa}}, \ and\ \bibinfo {author} {\bibfnamefont {Masahito}\
			\bibnamefont {Ueda}},\ }\bibfield  {title} {\enquote {\bibinfo {title}
			{{Topological Phases of Non-Hermitian Systems}},}\ }\href {\doibase
		10.1103/physrevx.8.031079} {\bibfield  {journal} {\bibinfo  {journal}
			{Physical Review X}\ }\textbf {\bibinfo {volume} {8}},\ \bibinfo {pages}
		{031079} (\bibinfo {year} {2018})}\BibitemShut {NoStop}%
	\bibitem [{\citenamefont {Jin}\ and\ \citenamefont {Song}(2019)}]{Jin2019}%
	\BibitemOpen
	\bibfield  {author} {\bibinfo {author} {\bibfnamefont {L.}~\bibnamefont
			{Jin}}\ and\ \bibinfo {author} {\bibfnamefont {Z.}~\bibnamefont {Song}},\
	}\bibfield  {title} {\enquote {\bibinfo {title} {{Bulk-boundary
					correspondence in a non-Hermitian system in one dimension with chiral
					inversion symmetry}},}\ }\href {\doibase 10.1103/physrevb.99.081103}
	{\bibfield  {journal} {\bibinfo  {journal} {Physical Review B}\ }\textbf
		{\bibinfo {volume} {99}},\ \bibinfo {pages} {081103} (\bibinfo {year}
		{2019})}\BibitemShut {NoStop}%
	\bibitem [{\citenamefont {Zhang}\ \emph {et~al.}(2019)\citenamefont {Zhang},
		\citenamefont {Wang},\ and\ \citenamefont {Song}}]{Zhang2019}%
	\BibitemOpen
	\bibfield  {author} {\bibinfo {author} {\bibfnamefont {K.~L.}\ \bibnamefont
			{Zhang}}, \bibinfo {author} {\bibfnamefont {P.}~\bibnamefont {Wang}}, \ and\
		\bibinfo {author} {\bibfnamefont {Z.}~\bibnamefont {Song}},\ }\bibfield
	{title} {\enquote {\bibinfo {title} {{Exceptional-point-induced lasing
					dynamics in a non-Hermitian Su-Schrieffer-Heeger model}},}\ }\href {\doibase
		10.1103/physreva.99.042111} {\bibfield  {journal} {\bibinfo  {journal}
			{Physical Review A}\ }\textbf {\bibinfo {volume} {99}},\ \bibinfo {pages}
		{042111} (\bibinfo {year} {2019})}\BibitemShut {NoStop}%
	\bibitem [{\citenamefont {Shi}\ \emph {et~al.}(2005)\citenamefont {Shi},
		\citenamefont {Li}, \citenamefont {Song},\ and\ \citenamefont
		{Sun}}]{Shi2005}%
	\BibitemOpen
	\bibfield  {author} {\bibinfo {author} {\bibfnamefont {Tao}\ \bibnamefont
			{Shi}}, \bibinfo {author} {\bibfnamefont {Ying}\ \bibnamefont {Li}}, \bibinfo
		{author} {\bibfnamefont {Zhi}\ \bibnamefont {Song}}, \ and\ \bibinfo {author}
		{\bibfnamefont {Chang-Pu}\ \bibnamefont {Sun}},\ }\bibfield  {title}
	{\enquote {\bibinfo {title} {Quantum-state transfer via the ferromagnetic
				chain in a spatially modulated field},}\ }\href {\doibase
		10.1103/physreva.71.032309} {\bibfield  {journal} {\bibinfo  {journal}
			{Physical Review A}\ }\textbf {\bibinfo {volume} {71}},\ \bibinfo {pages}
		{032309} (\bibinfo {year} {2005})}\BibitemShut {NoStop}%
	\bibitem [{\citenamefont {Tao}\ \emph {et~al.}(2005)\citenamefont {Tao},
		\citenamefont {Bing},\ and\ \citenamefont {Zhi}}]{Tao2005}%
	\BibitemOpen
	\bibfield  {author} {\bibinfo {author} {\bibfnamefont {Shi}\ \bibnamefont
			{Tao}}, \bibinfo {author} {\bibfnamefont {Chen}\ \bibnamefont {Bing}}, \ and\
		\bibinfo {author} {\bibfnamefont {Song}\ \bibnamefont {Zhi}},\ }\bibfield
	{title} {\enquote {\bibinfo {title} {{On Harmonic Approximation for Large
					Josephson Junction Coupling Charge Qubits}},}\ }\href {\doibase
		10.1088/0253-6102/43/5/006} {\bibfield  {journal} {\bibinfo  {journal}
			{Communications in Theoretical Physics}\ }\textbf {\bibinfo {volume} {43}},\
		\bibinfo {pages} {795--798} (\bibinfo {year} {2005})}\BibitemShut {NoStop}%
	{\bibitem [{\citenamefont {Scholtz}\ \emph {et~al.}(1992)\citenamefont
			{Scholtz}, \citenamefont {Geyer},\ and\ \citenamefont {Hahne}}]{Scholtz1992}%
		\BibitemOpen
		\bibfield  {author} {\bibinfo {author} {\bibfnamefont {F.G.}\ \bibnamefont
				{Scholtz}}, \bibinfo {author} {\bibfnamefont {H.B.}\ \bibnamefont {Geyer}}, \
			and\ \bibinfo {author} {\bibfnamefont {F.J.W.}\ \bibnamefont {Hahne}},\
		}\bibfield  {title} {\enquote {\bibinfo {title} {{Quasi-Hermitian operators
						in quantum mechanics and the variational principle}},}\ }\href {\doibase
			10.1016/0003-4916(92)90284-s} {\bibfield  {journal} {\bibinfo  {journal}
				{Annals of Physics}\ }\textbf {\bibinfo {volume} {213}},\ \bibinfo {pages}
			{74--101} (\bibinfo {year} {1992})}\BibitemShut {NoStop}}%
		{\bibitem [{\citenamefont {Brody}(2013)}]{Brody2013}%
		\BibitemOpen
		\bibfield  {author} {\bibinfo {author} {\bibfnamefont {Dorje~C}\ \bibnamefont
				{Brody}},\ }\bibfield  {title} {\enquote {\bibinfo {title} {Biorthogonal
					quantum mechanics},}\ }\href {\doibase 10.1088/1751-8113/47/3/035305}
		{\bibfield  {journal} {\bibinfo  {journal} {Journal of Physics A:
					Mathematical and Theoretical}\ }\textbf {\bibinfo {volume} {47}},\ \bibinfo
			{pages} {035305} (\bibinfo {year} {2013})}\BibitemShut {NoStop}}%
	\bibitem [{\citenamefont {Zhang}\ \emph {et~al.}(2024)\citenamefont {Zhang},
		\citenamefont {Zhang},\ and\ \citenamefont {Song}}]{Zhang2024}%
	\BibitemOpen
	\bibfield  {author} {\bibinfo {author} {\bibfnamefont {H.~P.}\ \bibnamefont
			{Zhang}}, \bibinfo {author} {\bibfnamefont {K.~L.}\ \bibnamefont {Zhang}}, \
		and\ \bibinfo {author} {\bibfnamefont {Z.}~\bibnamefont {Song}},\ }\bibfield
	{title} {\enquote {\bibinfo {title} {{Dynamics of non-Hermitian Floquet
					Wannier-Stark system}},}\ }\href@noop {} {\  (\bibinfo {year} {2024})},\
	\Eprint {http://arxiv.org/abs/2401.13286} {arXiv:2401.13286 [quant-ph]}
	\BibitemShut {NoStop}%
\end{thebibliography}
%

\end{document}